# Initial transient stage of pin-to-pin nanosecond repetitively pulsed discharges in air

Xingxing Wang, Adam Patel, Alexey Shashurin

## Abstract

In this work, evolution of parameters of nanosecond repetitively pulsed (NRP) discharges in pin-to-pin configuration in air was studied during transient stage of initial twenty discharge pulses. Gas and plasma parameters in the discharge gap were measured using coherent microwave scattering (CMS), optical emission spectroscopy (OES) and laser Rayleigh scattering (LRS) for NRP discharges at repetition frequencies of 1, 10 and 100 kHz. Memory effects (when perturbations induced by the previous discharge pulse would not decay fully till the subsequent pulse) were detected for the repetition frequencies of 10 and 100 kHz. For 10 kHz NRP discharge, the discharge parameters experienced significant change after the first pulse and continued to substantially fluctuate between the subsequent pulses due to rapid evolution of gas density and temperature during the 100 μs inter-pulse time caused by intense redistribution of the flow field in the gap on that time scale. For 100 kHz NRP discharge, the discharge pulse parameters reached a new steady-state at about five pulses after initiation. This new steady-state was associated with well-reproducible parameters between the discharge pulses and substantial reduction of breakdown voltage, discharge pulse energy, and electron number density in comparison with the first discharge pulse.

# I. Introduction

Nanosecond discharges (ns-discharges) have drawn great interest due to wide variety of potential applications. The ns-scale rise time of high voltage driving pulse allows the voltage to quickly increase beyond the breakdown voltage threshold which enables efficient energy deposition into an interelectrode gas volume and production of active species.[1] Utilization of series of ns-discharges applied at repetition frequencies in kHz range refers to a nanosecond repetitively pulsed (NRP) discharge. The NRP discharges found applications in plasma-assisted combustion owing to generation of gaseous species including O, H, OH, etc.[2–4] Additionally, NRP discharges can be used for aerodynamic flow control applications (plasma actuators) which utilize NRP-driven energy deposition into gas in vicinity of airfoil surface to delay flow separation and reduce drag.[5,6] NRP discharges have been also utilized in many other fields including medicine, nanotechnology, material processing, sterilization, etc.[7]

In our previous work,[8] ns-discharge in air in pin-to-pin configuration with 5 mm gap operating in the single-pulse mode was studied. Two discharge regimes were observed (spark and corona) and characterized. In spark regime, about 5 mJ nanosecond discharge pulse produced plasmas with electron number density of approximately $5\times10^{15}$ cm$^{-3}$ and characteristic decay time of 150-200 ns. Gas temperature was measured to be 3500 K at 10 μs after the discharge by optical emission spectroscopy (OES), and gas density reduced to a minimum of 30% of ambient value 5 μs after the discharge as detected by laser Rayleigh scattering. Both gas temperature and gas number density recover to ambient conditions within about 1 ms after the discharge pulse. Therefore, one should expect the onset of memory effects (when perturbations induced by the previous discharge pulse would not decay fully till the subsequent pulse) at repetition frequency higher than 1 kHz.

NRP pin-to-pin discharges operating at repetition frequencies in kHz range have been studied by many other groups. In the works by Pai et. al. and Rusterholtz et. al.,[9–11] NRP pin-to-pin discharges at 10-30 kHz were studied in a preheated air at 1000 K where the gap distance between electrodes was kept at 4-5 mm. With an energy input of <1 mJ, gas temperature was measured to be up to 5000 K, and the electron number density was on the order of $10^{15}$ cm$^{-3}$. In the work by Horst et al.,[12] NRP pin-to-pin discharge was studied in $N_2$ and 0.9% humidified $N_2$. The discharges were operating at a repetition frequency of 1 kHz at atmospheric conditions with a gap of 2 mm. The gas temperature was measured to be up to 750 K at the moment 1 μs after the nanosecond pulse, and the electron number density was reported to be on the order of $10^{16}$ cm$^{-3}$ when the energy input per pulse is around 1 mJ.

Therefore, air-based NRP discharges in pin-to-pin configuration operating in kHz frequency range were studied previously. However, parameters of these discharges were mostly studied on a later stage well after initiation of sequence of the discharge pulses while transition of parameters from pulse #1 to the later pulses was not thoroughly studied. In this work, transient dynamics of gas and plasma parameters for first twenty pulses after the NRP discharge initiation for different discharge repetition frequencies was studied.

## II. Experimental setup and Methodology

The experimental setup utilized HV nanosecond pulse (25 kV peak, 90 ns width, repetition frequency up to 400 kHz) supplied by a ns-pulser (Eagle Harbor NSP-3300-20-F) to initiate NRP discharges and pin-to-pin discharge electrodes separated 5 mm apart. Voltage (*V*) and current (*I*) measurements were conducted by conventional voltage probes (Tektronix P6015A) and current transformer (Bergoz FCT-028-0.5-WB). Gas temperature $T_{gas}$ (= rotational temperature $T_{rot}$)[13] was measured by optical emission spectroscopy (OES) of nitrogen second positive system using Princeton Instruments SP-2750i spectrometer and Princeton Instruments PI-MAX 1024i ICCD camera. Gas density ($n_g$) measurements were conducted by laser Rayleigh scattering (LRS) using nanosecond laser at 532 nm supplied by EKSPLA NT342. Finally, electron number density ($n_e$) was measured using coherent microwave scattering (CMS) combined with LRS. The experimental setup was identical to one used in our previous work.[8]

Gas temperature $T_{gas}$ and bulk gas density $n_g$ were recorded at pulses # 1, 2, 3, 4, 5, 10, and 20 of the durst. $T_{gas}$ on each pulse was recorded at the beginning of the ns-pulse representing the initial condition of temperature for the corresponding discharge. The timing of the ICCD camera gate was controlled via the gate delay setting in Lightfield software. Two measurements of $n_g$ were recorded for each discharge pulse by sending the laser pulses at desired moments controlled by an external waveform generator. Specifically, the laser pulses were sent right before the initiation of the discharge pulse for the measurement of initial condition of $n_g$ of each discharge ($n_{g0}$) and 5 μs after the initiation of the discharge pulse which is expected to be representative for the lowest gas number density after discharge ($n_{g,min}$).[8,14] In order to reach adequate signal-to-noise ratio, 100 accumulations were taken for the OES measurement and 10 accumulations were used for the LRS measurement.

For the CMS system, $n_e$ was determined from the output CMS signal $U_s$ using following expression:[15] $U_s = \frac{Ae^2 n_e V}{m\sqrt{\omega^2 + \nu_m^2}}$, where $A$ – the CMS system calibration factor (determined using calibration with dielectric scatterers), V - plasma volume (determined using ICCD photography), $\nu_m$- electron-gas collisional frequency (determined based on LRS measurements), $\omega$- microwave frequency, and $m$ and $e$ – electron mass and charge. More details on $n_e$-measurements by means of combination of CMS and LRS techniques can be found elsewhere.[8,15]

## III. Results and discussion

The temporal evolution of discharge parameters for a burst of 20 pulses applied at repetition rate of $f = 1$ kHz are presented in Figure 1 including discharge voltage ($V_d$), discharge current ($I_d$), pulse energy ($E_{pulse}$, calculated by integrating the product of voltage and current over pulse duration), and pre-pulse gas temperature ($T_{g0}$) and density ($n_{g0}$) measured right before the initiation of corresponding discharge pulse. Note that Figure 1(a) and Figure 1(b) intentionally present temporally unresolved dynamics for $V_d$ and $I_d$ to trace evolution of a peak values of corresponding properties in the burst, namely, breakdown voltage $V_{br}$ and peak discharge current $I_{peak}$. Additionally, $n_e$ plotted in Figure 1(f) was determined from Eq. 1 using measured microwave scattering signal $U_s$, plasma volume of V=2.5×10$^{-4}$ cm$^3$ (a cylinder with height of 5 mm and diameter of 250 µm) determined using ICCD camera, and following approximation for collisional frequency $\nu_m = \frac{1.42 \times 10^{12}}{2.5 \times 10^{19}} n_{g0} (\text{cm}^{-3}), \text{s}^{-1}$ used in combination with measured $n_{g0}$ dynamics across pulses in the burst.[15] The utilized approximations for V and $\nu_m$ are valid during the ns-discharge pulse while deviate on the decay stage;[8,14] therefore, Figure 1(f) allows to trace evolution of peak electron number density $n_{e,peak}$ along the pulses in the burst while details of electron decay are intentionally unresolved in time.

One can see in the Figure 1 that discharge parameters were similar between different pulses in the burst and experienced fairly small variations (on the order of ±6%, ±12%, ±2%, ±6%, ±3%, ±10% for $V_{br}$, $I_{peak}$, $E_{pulse}$, $T_{g0}$, $n_{g0}$, $n_{e,peak}$, respectively) around corresponding steady-state values

($V_{br}$=19 kV, $I_{peak}$=16 A, $E_{pulse}$=5.1 mJ,, $T_{g0}$=300 K, $n_{g0}$=2.5×10$^{19}$ cm$^{-3}$, and $n_{e,peak}$=1.5×10$^{16}$ cm$^{-3}$). These variations between the discharge pulses are associated with natural irreproducibility of the streamer breakdown and microscale changes of the cathode morphology due to cathode spots presence on the later stage of the discharge pulse.[8] Temporally-resolved discharge voltage and current waveforms were also consistent across the pulses in the burst, and typical evolutions of $V_d$ and $I_d$ are shown in Figure 2 (for pulse #10). Thus, it might be concluded that each discharge pulse in the burst closely resembles pulse #1 which indicates that no memory effects between the pulses are present. This conclusion is consistent with our recent findings that gas parameters ($n_g$ and $T_g$) in the gap fully recover to their initial unperturbed pre-discharge values on times ~1 ms after the ns-pulse.[8] Therefore, each subsequent discharge pulse in the burst occurs in the gas of essentially same properties (300 K, 2.5×10$^{19}$ cm$^{-3}$) yielding similar discharge parameters for all pulses.

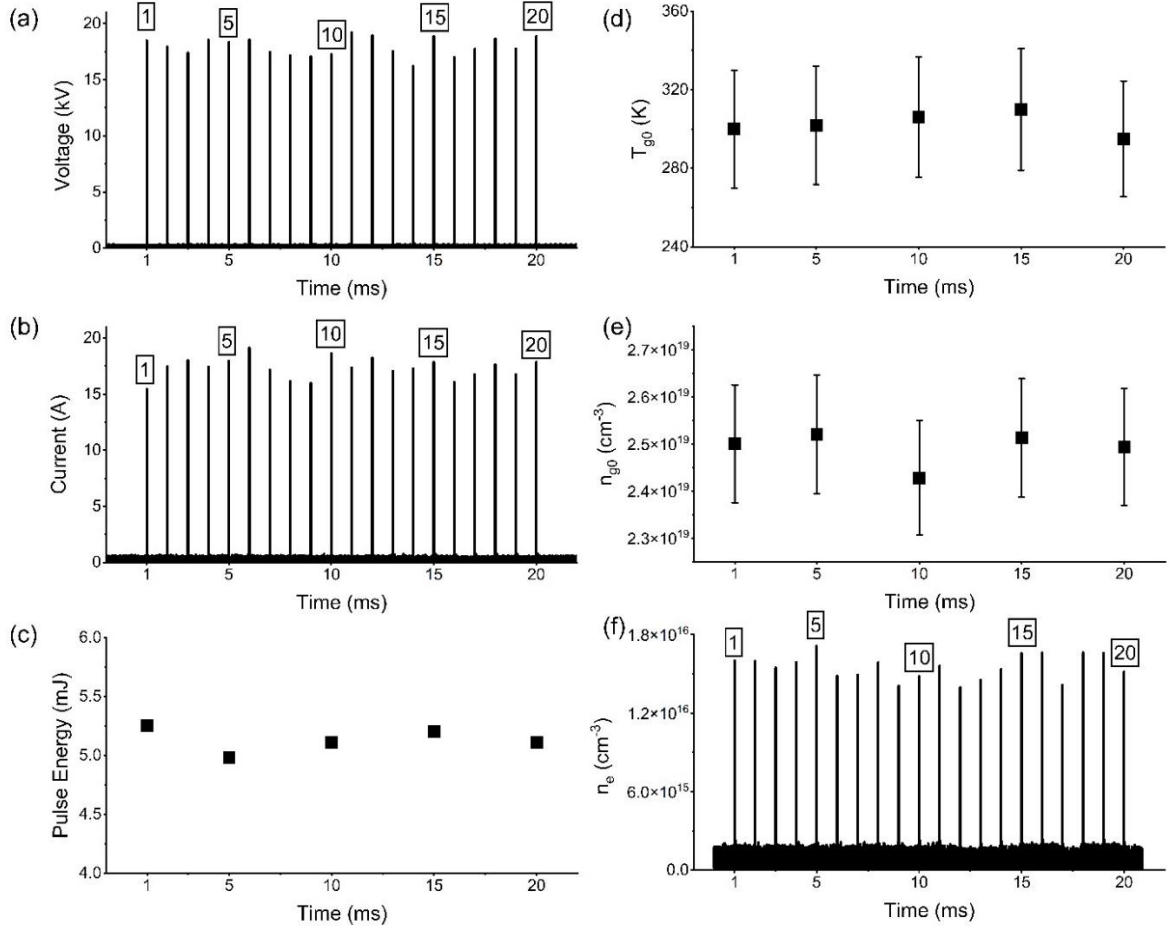

**Figure 1. Temporal evolution of parameters for NRP discharge at $f$ = 1 kHz. (a) voltage ($V_d$); (b) current ($I_d$); (c) discharge pulse energy ($E_{pulse}$); (d) gas temperature $T_{g0}$ right before discharge pulse initiation (for pulses #1,5,10, 15, 20); (d) gas number density $n_{g0}$ right before discharge pulse initiation (for pulses #1,5,10, 15, 20); (f) electron number density $n_e$.**

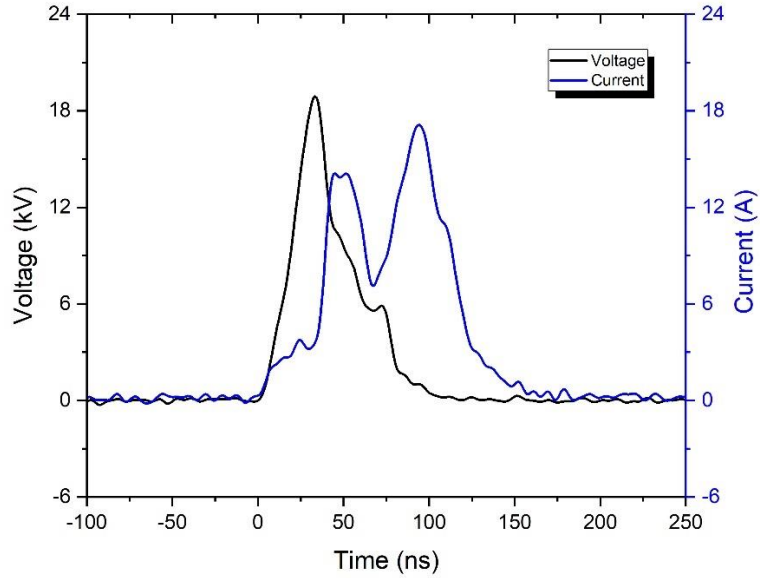

**Figure 2. Typical temporal evolution of discharge voltage (black) and discharge current (blue) for pulses in the burst ($f$ = 1 kHz, pulse #10).**

Evolution of the discharge parameters for a burst of twenty pulses at the repetition rate of $f$ = 10 kHz is presented in Figure 3 (note that $n_e$ was determined using same approximations for V=2.5×10$^{-4}$ cm$^3$ and $v_m = \frac{1.42 \times 10^{12}}{2.5 \times 10^{19}} n_{g0}(\text{cm}^{-3}), \text{s}^{-1}$ as above for $f$ = 1 kHz). One can see that discharge parameters of the first pulse substantially differ from that of consecutive pulses. Specifically, $V_{br}$, $E_{pulse}$, $n_{e,peak}$ decreased from 18 kV, 5.2 mJ, 1.5×10$^{16}$ cm$^{-3}$ for pulse #1 to 5-14.5 kV, 2.6-3.8 mJ, 2.5-5×10$^{15}$ cm$^{-3}$ for pulses #2-20. This change of parameters can be explained by a memory effect; namely, gas number density and temperature do not recover to their initial unperturbed values of 300 K and 2.5×10$^{19}$ cm$^{-3}$ by the time of the second pulse arrival. Indeed, Figure 3 (c) supports that gas temperature and number density before the second pulse were substantially different, namely, $T_{g0}$=1600 K and $n_{g0}$ =1.1×10$^{19}$ cm$^{-3}$. Finally, the memory effects were preserved for the entire duration of the burst given that $T_g$ and $n_g$ never recover to their initial unperturbed values (300 K, 2.5×10$^{19}$ cm$^{-3}$) as one can see in Figure 3(d) and (e).

Starting from pulse #2, the discharge pulse parameters shown in Figure 3 settle on a new steady-state values of $V_{br}$=9.5 kV, $I_{peak}$=20 A, $E_{pulse}$=3.2 mJ, $T_{g0}$=1100 K, $n_{g0}$=1.0×10$^{19}$ cm$^{-3}$, $n_{e,peak}$=3.7×10$^{15}$cm$^{-3}$ which are different from that at $f$ =1 kHz. Additionally, these new steady-state values are associated with more substantial variability between the pulses compared to the case of $f$ =1 kHz; namely, variations of $V_{br}$, $I_{peak}$, $E_{pulse}$, $T_{g0}$, $n_{g0}$, $n_{e,peak}$ were ±49%, ±13%, ±14%, ±10%, ±13% and ±35%, respectively. This larger variability of the discharge pulse parameters refers to the fact that moment of subsequent pulse initiation coincides with the region of rapid temporal change of $T_g$ and $n_g$ which is associated with intense redistribution of the flow field in the gap on a 100 μs inter-pulse time scale.[16] Indeed, our recent measurements confirm a very rapid change of $T_g$ and $n_g$ during 100 μs after the first discharge pulse (about 60% drop for $T_g$ and 30% increase for and $n_g$, respectively).[8] Additionally, minimal number density for each discharge pulse, $n_{g,min}$, (which establishes on a μs-timescale after the discharge pulse)[8] is presented in Figure 3(e) to demonstrate variation of $n_g$ in subsequent discharge pulses (about 12-34 %). Thus, substantial variation of $T_{g0}$ and $n_{g0}$ prior to each discharge pulse in the burst is expected, which causes relatively large variation of the discharge parameters observed in the experiments as shown in Figure 3. Note that substantially lower $E_{pulse}$ for pulses #2-20 (2.6-3.8 mJ) compared to pulse #1 (5.2 mJ) indicate that pulser was better matched with the load (discharge gap filled with plasma) on the first pulse (impedance immediately after the breakdown was ~1 kΩ as one can estimate from Figure 2) rather than for the subsequent pulses (impedance immediately after the breakdown was ~100-150 Ω according to Figure 4). Correspondingly, lower $E_{pulse}$ for pulses #2-20 caused lower peak electron number density observed in these pulses.

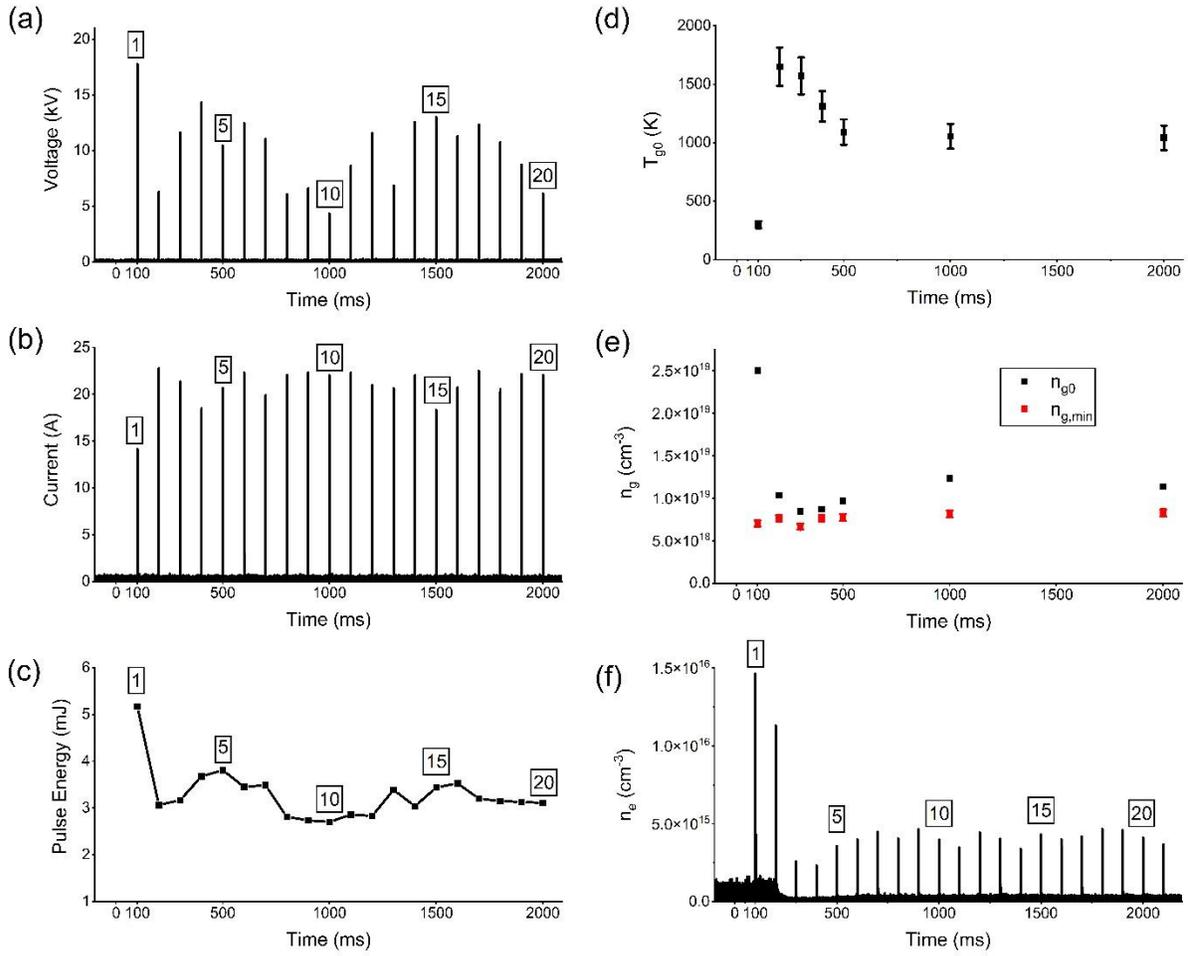

**Figure 3.** Temporal evolution of parameters for NRP discharge at $f = 10$ kHz. (a) voltage ($V_d$); (b) current ($I_d$); (c) discharge pulse energy ($E_{pulse}$); (d) gas temperature $T_{g0}$ right before discharge pulse initiation for pulses #1, 2, 3, 4, 5, 10, 20; (e) gas number density $n_{g0}$ right before discharge pulse initiation (black) and minimum gas density $n_{g,min}$ after each discharge (red) for pulses #1, 2, 3, 4, 5, 10, 20; (f) electron number density $n_e$.

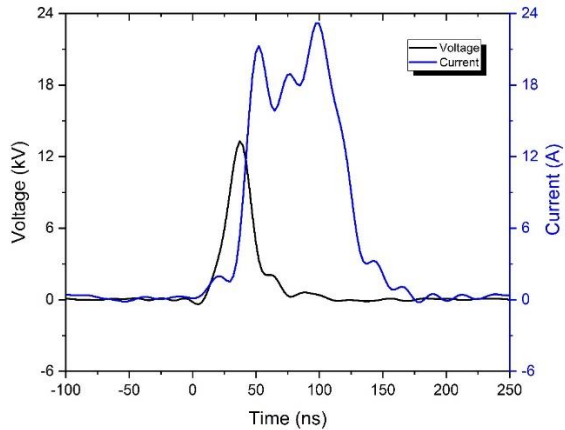

**Figure 4.** Typical temporal evolution of discharge voltage (black) and discharge current (blue) for pulses in the burst ($f = 10$ kHz, pulse #10).

The evolution of discharge parameters for a burst of 20 pulses at repetition rate of $f = 100$ kHz is presented in Figure 5. $n_e$ was determined from CMS measurements using same approximation for V $=2.5\times10^{-4}$ cm$^3$, while $\nu_m$-approximation in the form $\nu_m = \frac{1.42\times10^{12}}{2.5\times10^{19}} n_{g0}(\text{cm}^{-3})$, s$^{-1}$ based on measured $n_{g0}$-values was applied starting pulse #2 only (given that $n_g$ experienced very minor variations of 5-10% starting pulse #2). For pulse #1, the temporal evolution of $\nu_m$ was used for $n_e$ determination based on temporally-resolved measurements of $n_g$ by laser Rayleigh scattering reported in Refs. [8,14].

Similar to the 10 kHz case, memory effects were obviously present causing variation of the discharge pulse parameters from that of the first pulse. Steady-state discharge pulse parameters at 100 kHz were reached at about pulse #5 and were associated with lower variability compared to the 10 kHz case ($V_{br}$=6 kV, $I_{peak}$=23 A, $E_{pulse}$=1.4 mJ, $T_{g0}$=5000 K, $n_{g0}$=0.25×10$^{19}$ cm$^{-3}$, $n_{e,peak}$=1.5×10$^{15}$ cm$^{-3}$ with corresponding variations of ±12%, ±5%, ±6%, ±2%, ±20%, ±3%, respectively). Smaller variability in the steady-state is associated with short time (10 μs) between the pulses so that gas properties $T_g$ and $n_g$ did not change significantly on that inter-pulse timescale. This is supported by $n_g$-plot in Figure 5(e) showing very minor variation between $n_{g0}$ and $n_{g,min}$ (about 5-10 %) starting pulse #2. Even lower pulse energy $E_{pulse}$ =1.4 mJ in the steady-state of the 100 kHz NRP discharge compared to the 10 kHz counterpart (3.2 mJ) indicate even greater pulser's mismatch with the discharge gap load (impedance immediately after breakdown was ~75 Ω according to Figure 6). This smaller steady-state pulse energy for the 100 kHz NRP discharge correspondingly led to smaller electron number density $n_e$=1.5×10$^{15}$ cm$^{-3}$.

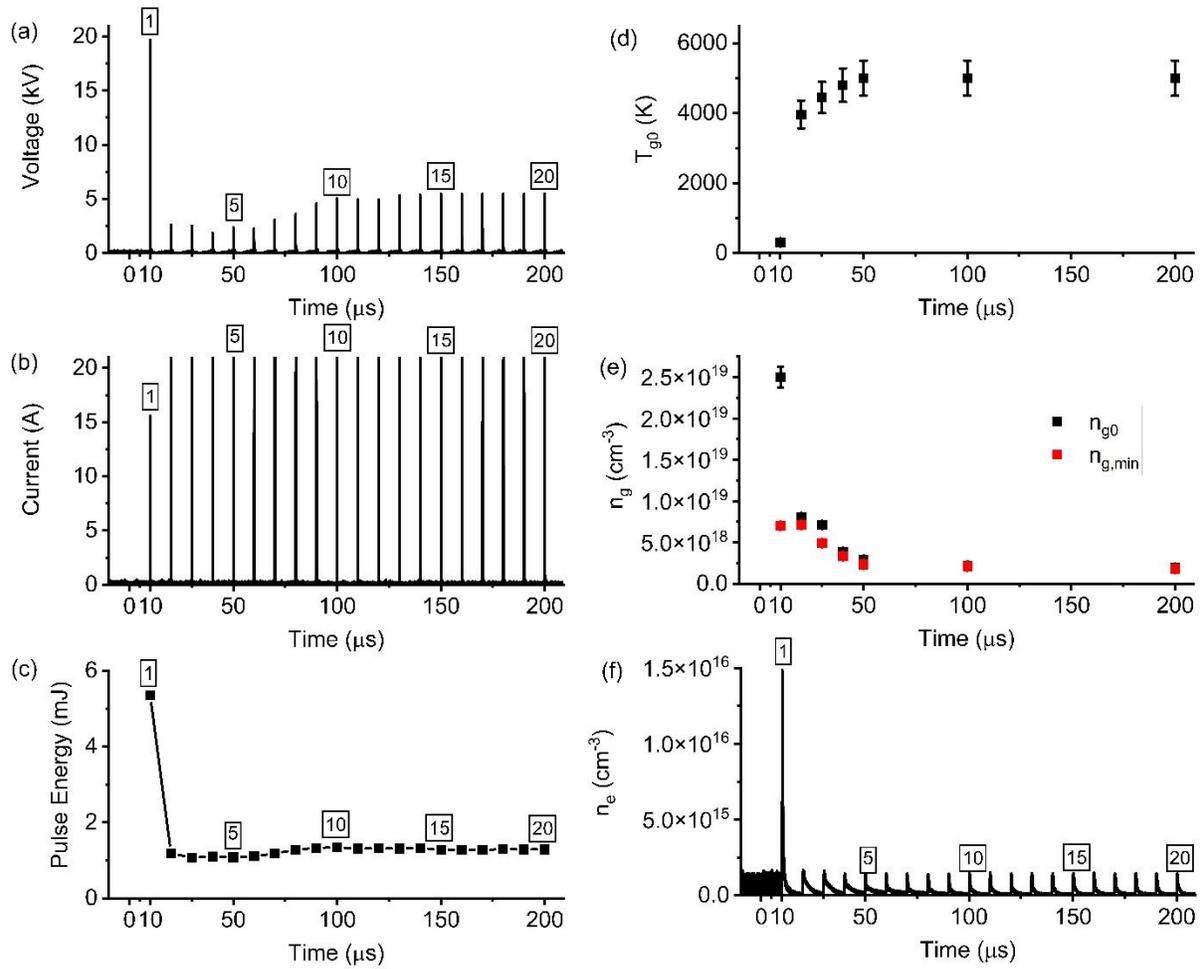

**Figure 5.** Temporal evolution of parameters for NRP discharge at $f$ = 100 kHz. (a) voltage ($V_d$); (b) current ($I_d$); (c) discharge pulse energy ($E_{pulse}$); (d) gas temperature $T_{g0}$ right before discharge pulse initiation for pulses #1, 2, 3, 4, 5, 10, 20; (e) gas number density $n_{g0}$ right before discharge pulse initiation (black) and minimum gas density $n_{g,min}$ after each discharge (red) for pulses #1, 2, 3, 4, 5, 10, 20; (f) electron number density $n_e$.

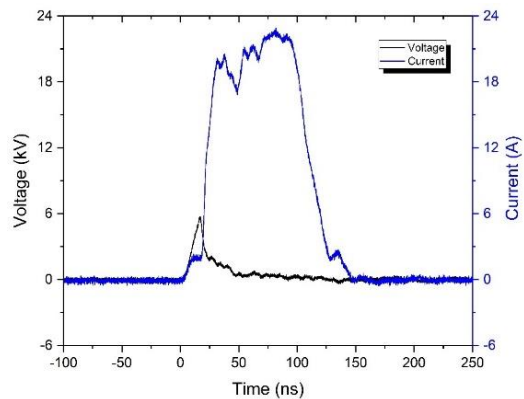

**Figure 6.** Typical temporal evolution of discharge voltage (black) and discharge current (blue) for pulses in the burst ($f$ = 100 kHz, pulse #10).

Figure 5 (f) allows direct comparison of the electron decay of single-pulse ns-discharge (based on pulse #1) and steady-state of the 100 kHz NRP discharge (after pulse #5). For the single-pulse ns-discharge, the characteristic decay time was 150-200 ns (corresponding decay rate of 0.5-0.7×$10^7$ $s^{-1}$) while for the steady-state of the 100 kHz NRP discharge it increased to 400-700 ns (decay rate of 0.14-0.25×$10^7$ $s^{-1}$). In general, electron decay is driven by dissociative recombination ($X_2^+ + e \rightarrow X + X$, where $X$ is $N$ or $O$) and three-body attachment to oxygen ($e + O_2 + X_2 \rightarrow O_2^- + X_2$, where $X$ is $N$ or $O$) governed by the following equation: $\frac{dn_e}{dt} = -\beta n_e^2 - \nu n_e$.[8] We have numerically simulated this equation using electron temperatures in the range 0.3-3 eV, dissociative recombination rate for oxygen and nitrogen $\beta = 2 \times 10^{-7} \times \sqrt{\frac{300}{T_e}}$ [$cm^3 s^{-1}$], and three-body attachment rate to oxygen $\nu = k_1 n_{O_2}^2 + k_2 n_{O_2} n_{N_2}$, where $k_1 = 1.4 \times 10^{-29} \times \frac{300}{T_e} \times \exp\left(-\frac{600}{T_{gas}}\right) \times \exp\left(\frac{700 \times (T_e - T_{gas})}{T_e \times T_{gas}}\right)$ [$cm^6 s^{-1}$] and $k_2 = 1.07 \times 10^{-31} \times \left(\frac{300}{T_e}\right)^2 \times \exp\left(-\frac{70}{T_{gas}}\right) \times \exp\left(\frac{1500 \times (T_e - T_{gas})}{T_e \times T_{gas}}\right)$ [$cm^6 s^{-1}$].[17,18] Contribution of dissociative recombination was dominant in the initial portion of the electron decay for both single-pulse ns-discharge and 100 kHz NRP discharge. Numerical simulations yield substantially faster decay times (1-2 orders of magnitude) in comparison with the experimentally measured ones; namely, $T_e$= 0.3, 1 and 3 eV yield decay times 1, 2, and 4 ns for the single-pulse discharge, and 8, 15, and 26 ns for 100 kHz NRP discharge in steady-state. The observed discrepancy can be potentially attributed to inapplicability of rate coefficients available in literature to conditions of NRP discharges (due to high gas temperatures, presence of excited ions rather than in ground state)[8] and to additional ionization of gas after ns-pulse in collisions of electrons with excited gas particles.

## Conclusions

Initial transient stage of NRP pin-to-pin discharges is associated with first several discharge pulses, e.g., first five (5) pulses for the case of few mJ nanosecond discharge pulses, 5 mm discharge gap, and 100 kHz repetition frequency used in this work. Conditions of the gas in the discharge gap significantly change during these initial pulses leading to manifold reduction of gas number density and increase of gas temperature. This, in turn, affects impedance of the interelectrode gap filled with plasma during the discharge and, correspondingly, changes energy deposition into the gap in subsequent discharge pulses. In general, electron decay after individual discharge pulses is governed by contributions of dissociative recombination and electron attachment to oxygen, while dissociative recombination is expected to dominate initial portion of the decay. Relatively slow measured electron decay can be potentially attributed to inapplicability of reaction rate coefficients available in literature to conditions of NRP discharges and to additional ionization of gas after ns discharge pulses in collisions of electrons with excited gas particles.

## Acknowledgment

Authors would like to thank N. A. Popov, M. N. Shneider, and S. Bane for valuable discussions. This work was supported by the U.S. Department of Energy (Grant No. DE-SC0018156) and partially by the National Science Foundation (Grant No. 1903415).

## Data availability

The data that supports the findings in this study are available from the corresponding author upon reasonable request.